\documentclass[12pt]{article}
\usepackage{epsfig}
\begin{document}
\begin{center}
{\Large {\bf A model for fermion mass generation in Technicolor theory}


{
\vspace{1cm}
{ M.A.~Zubkov }\\
\vspace{.5cm} {\it  ITEP, B.Cheremushkinskaya 25, Moscow, 117259, Russia } }}
\end{center}

\begin{abstract}
We consider the $SU(N_{TC})$ Farhi - Susskind Technicolor model, in which
$SU(2)$ doublets of technifermions are right-handed while $SU(2)$ singlets of
technifermions are left-handed. We add coupling of fermions and technifermions
to $SU(N_{TC})$ fundamental massive scalar fields. Due to this coupling the
transitions between both types of fermions occur. Therefore the Standard Model
fermions acquire masses.
\end{abstract}

\section{Introduction}

It is commonly believed that due to the so-called Hierarchy problem the
Standard Model (SM) does not work at the energies above $1$ Tev\cite{TEV}. In
addition the indications are found that the lattice Weinberg - Salam Model
cannot have in principle the ultraviolet cutoff larger than about $1$
Tev\cite{Z2010,VZ2008}. That's why it is interesting to consider the models, in
which SM Higgs field appears as a composite field. One of the most popular
schemes is given within the Technicolor theory (TC), which provides dynamical
Electroweak symmetry breaking due to the spontaneous chiral symmetry breakdown
and the formation of the condensate for the
technifermions\cite{Technicolor,Technicolor_,Technicolor__}. However,
Technicolor theory alone cannot provide Standard Model fermions with realistic
masses. Usually, in order to make SM fermions massive the so-called Extended
Technicolor (ETC) is
introduced\cite{ExtendedTechnicolor,ExtendedTechnicolor_,ExtendedTechnicolor__,ExtendedTechnicolor___,ExtendedTechnicolor____},
which provides transition between SM fermions and technifermions. Due to this
transition SM fermions are coupled to the condensate of technifermions and
become massive. However, this scheme has several difficulties. Dangerous FCNC
as well as contributions to Electroweak polarization operators appear. Thus,
for example, the realistic masses for SM fermions cannot be generated without
appearance of the physical effects excluded by the present measurements.
Probably, a possible way to overcome these difficulties is related to the
consideration of the so-called walking
technicolor\cite{walking,minimal_walking,minimal_walking_,minimal_walking__}
which, however, helps reasonably well for all fermions except for the $t$
quark. Another problem related to Extended Technicolor is that a lot of other
Higgs fields are to appear in order to break ETC gauge group down to TC gauge
group. The origin of these fields and the pattern of the correspondent
breakdown is not considered in sufficient details at the present moment.

In the present paper we suggest an alternative scenario of SM fermions mass
formation. Namely, we consider the analogue of $SU(N_{TC})$ Farhi-Susskind
Technicolor model\cite{FS}, in which $SU(2)$ doublets are right-handed while
$SU(2)$ singlets are left - handed. With this change made the Technicolor model
leads to the same masses of $W$ and $Z$ bosons as the original one. At the same
time we are able to introduce an additional coupling of SM fermions and
technifermions to $SU(N_{TC})$ massive scalars. The action for these scalars
does not contain dangerous $\phi^4$ interactions. Therefore, problems specific
for the Higgs field of the SM are avoided. We assume the masses of the
mentioned scalars are essentially higher than the Technicolor scale. Therefore,
integrating out these fields we obtain the low energy effective four - fermion
interactions that allow coupling of SM fermions to technifermion condensates.
Thus, masses of the SM fermions appear in a complete analogy with ETC models.
However, the given approach has several advances. In particular, additional
FCNC do not appear at all. It is worth mentioning that in our approach neutrino
mass matrix is of Dirac - type. Therefore, some other physics is to be added if
it is necessary to obtain Majorana - type neutrino mass matrix.

\section{A model for one generation of quarks}

Let us consider first the model that contains quarks and techniquarks of one
generation. We arrange them in the following $SU(2)$ doublets and singlets.

Quarks:
\begin{equation}
l^{\omega i} = \frac{1-\gamma^5}{2} \left(\begin{array}{c}u^{\omega}\\
d^{\omega}\end{array}\right)=  \left(\begin{array}{c}u^{\omega}_L\\
d^{\omega}_L\end{array}\right)\, \, ; \, \, u^{\omega}_R =
\frac{1+\gamma^5}{2}u^{\omega} \,\, ; \, \, d^{\omega}_R =
\frac{1+\gamma^5}{2}d^{\omega} \label{q}
\end{equation}

Techniquarks:
\begin{equation}
R^{\omega a i} = \frac{1+\gamma^5}{2} \left(\begin{array}{c}U^{\omega a}\\
D^{\omega a}\end{array}\right)= \left(\begin{array}{c}U^{\omega a}_R\\
D^{\omega a}_R\end{array}\right)\, \, ; \, \, U^{\omega a}_L =
\frac{1-\gamma^5}{2}U^{\omega a} \,\, ; \, \, D^{\omega a}_L =
\frac{1-\gamma^5}{2}D^{\omega a} \label{Q}
\end{equation}
Here $a$ is $SU(N_{TC})$ index; $i$ is $SU(2)$ index; $\omega$ is color $SU(3)$
index. Hypercharge assignment for left - handed (right-handed) technifermions
is identical to that of the right - handed (left-handed) SM fermions. Contrary
to usual formulation of Technicolor the $SU(2)$ doublets of techniquarks are
right-handed while singlets are left - handed. This does not change, however,
the pattern of chiral symmetry breaking and the values of masses of $W$ and $Z$
bosons generated. (For general description of  how vacuum alignment works in
Technicolor theories see, for example, \cite{Align}. )

We also introduce $SU(N_{TC})$ fundamental scalar $H^a$. It is assumed that the
action for $H^a$ has the form:
\begin{equation}
S_H = \int (DH^+ DH + \frac{M^2}{4} H^2)d^4x
\end{equation}

We consider the action for massless quarks and Techniquarks with the additional
interaction term
\begin{equation}
S_I = \int \{(\bar{l}_{\omega i} R^{\omega ai} + \bar{u}_{R,\omega} U^{\omega
a}_L)H^+_a + (\bar{R}_{\omega ai} l^{\omega i} + \bar{U}_{L,\omega a}
u^{\omega}_R)H^a\}d^4x
\end{equation}

Technicolor provides chiral symmetry breaking. Therefore, Electroweak symmetry
is broken, $W$ and $Z$ bosons become massive and techniquarks are condensed:
\begin{eqnarray}
 <\bar{U}_{L,\omega a} R^{\rho a i}> & = & <\bar{R}_{\rho a i} U_L^{\omega a}> =\frac{1}{6}\delta^i_1 \delta^{\omega}_{\rho}<\bar{U} U>=-\frac{1}{6}\delta^i_1 \delta^{\omega}_{\rho}\Lambda_{TC}^3
\nonumber\\<\bar{D}_{L,\omega a} R^{\rho ai}> & = & <\bar{R}_{\rho ai}
D_L^{\omega a}>=\frac{1}{6}\delta_2^i \delta^{\omega}_{\rho}<\bar{D}
D>=-\frac{1}{6}\delta_2^i \delta^{\omega}_{\rho} \Lambda_{TC}^3
\end{eqnarray}
Here $\Lambda_{TC} \sim 1$ Tev is at the Technicolor scale.

We suppose $M >> \Lambda_{TC}$ and at low energies integration over $H$ leads
to effective four - fermion interaction term:
\begin{equation}
S_I = -\frac{1}{M^2}\int (\bar{l}_i R^{ai} + \bar{u}_R U^a_L)(\bar{R}_{ai}
l^{i} + \bar{U}_{L,a} u_R)d^4x
\end{equation}

Using Fierz rearrangement we rewrite this term in the following form:
\begin{eqnarray}
S_I &=& \frac{1}{M^2}\int \{-(\bar{l}_i R^{ai})(\bar{R}_{ai} l^{i})  -
(\bar{u}_R U^a_L)(\bar{U}_{L,a} u_R)  + \frac{1}{2}[(\bar{l}_{\omega i}
u^{\rho}_R)(\bar{U}_{L,\rho a} R^{\omega ai})\nonumber\\&&
 - \frac{1}{4}(\bar{l}_{\omega i}\gamma^{[\mu}\gamma^{\nu]}
u^{\rho}_R)(\bar{U}_{L,\rho a}\gamma_{[\mu} \gamma_{\nu]}R^{\omega ai})+
(c.c.)]\}d^4x\label{SI}
\end{eqnarray}
Here $(c.c.)$ means complex conjugate.

From (\ref{SI}) it follows that the mass term for $u$ - quark appears with the
mass
\begin{equation}
m_u = -\frac{1}{6M^2}<\bar{U}_{L,a} R^{a1}> =\frac{\Lambda^3_{TC}}{12M^2}
\end{equation}

\section{The model with all SM fermions}

Now let us turn to the whole Standard Model. We make the following notations:
\begin{eqnarray}
{\nu}^{b}& = & \left(\begin{array}{c} \nu \\
\nu_{\mu} \\
\nu_{\tau}\\\end{array}\right) \, ;\, u^{\omega b} =  \left(\begin{array}{ccc}  u^1 & u^2& u^3\\
c^1 &c^2 &c^3\\
t^1 &t^2 &t^3\\\end{array}\right) \nonumber\\
e^{b}& = & \left(\begin{array}{c} e \\
{\mu} \\
{\tau} \end{array}\right)\,;\,d^{\omega b} = \left(\begin{array}{ccc}  d^1 & d^2& d^3\\
s^1 &s^2 &s^3\\
b^1 &b^2 &b^3\end{array}\right)
\end{eqnarray}
Here index $\omega$ runs over $1,2,3$.   So the model contains the following
fields:

Quarks:
\begin{equation}
l^{\omega b i} = \frac{1-\gamma^5}{2} \left(\begin{array}{c}u^{\omega b}\\
d^{\omega b}\end{array}\right)=  \left(\begin{array}{c}u^{\omega b}_L\\
d^{\omega b}_L\end{array}\right)\, \, ; \, \, u^{\omega b}_R =
\frac{1+\gamma^5}{2}u^{\omega b} \,\, ; \, \, d^{\omega b}_R =
\frac{1+\gamma^5}{2}d^{\omega b} \label{q}
\end{equation}

Leptons:
\begin{equation}
l^{ b i} = \frac{1-\gamma^5}{2} \left(\begin{array}{c}\nu^{ b}\\
e^{b}\end{array}\right)=  \left(\begin{array}{c}\nu^{ b}_L\\
e^{b}_L\end{array}\right)\, \, ; \, \, \nu^{ b}_R = \frac{1+\gamma^5}{2}\nu^{
b} \,\, ; \, \, e^{ b}_R = \frac{1+\gamma^5}{2}e^{ b} \label{q}
\end{equation}

Techniquarks:
\begin{equation}
R^{\omega a  i} = \frac{1+\gamma^5}{2} \left(\begin{array}{c}U^{\omega a}\\
D^{\omega a}\end{array}\right)= \left(\begin{array}{c}U^{\omega a}_R\\
D^{\omega a i}_R\end{array}\right)\, \, ; \, \, U^{\omega a}_L =
\frac{1-\gamma^5}{2}U^{\omega a} \,\, ; \, \, D^{\omega a}_L =
\frac{1-\gamma^5}{2}D^{\omega a} \label{Q}
\end{equation}

Technileptons:
\begin{equation}
R^{ a i} = \frac{1+\gamma^5}{2} \left(\begin{array}{c}N^{ a}\\
E^{a}\end{array}\right)=  \left(\begin{array}{c}N^{a}_R\\
E^{a}_R\end{array}\right)\, \, ; \, \, N^{a}_L = \frac{1-\gamma^5}{2}N^{a} \,\,
; \, \, E^{a}_L = \frac{1-\gamma^5}{2}E^{a} \label{q}
\end{equation}

Here we have the only generation of technifermions. Index $a$ belongs to
$SU(N_{TC})$. Hypercharge assignment for left - handed (right-handed)
technifermions is identical to that of the right - handed (left-handed) SM
fermions.

Let us consider the following interaction term:
\begin{eqnarray}
&& S_I = \int \{\delta^{b}_{c} \bar{l}_{\omega b i} R^{\omega a i} +
\eta^{b}_{c} \bar{u}_{R,\omega,b} U^{\omega a}_L  +
\lambda^{b}_{c}\bar{d}_{R,\omega,b} D^{\omega a}_L\}H^{c}_a
d^4x\nonumber\\&&+\int \{\delta^{b}_{c} \bar{l}_{ b i} R^{ a i} +
\zeta^{b}_{c}\bar{\nu}_{R,b} N^{ a}_L +\gamma^{b}_{c}\bar{e}_{R,b} E^{
a}_L\}H^{c}_a d^4x\nonumber\\&&+ (c.c.)
\end{eqnarray}
Here $\eta$, $\lambda$, $\zeta$, and $\gamma$ are hermitian matrices of
coupling constants. We have introduced $3$ $SU(N_{TC})$ scalar fields $H^{c}$
with the common action
\begin{equation}
S_H = \int ( [DH_a^{c }]^+ DH_a^{c} + \frac{M^2}{4} [H_a^{c}]^+H_a^{c})d^4x
\end{equation}

Integrating out scalar fields we obtain low energy effective four - fermion
interactions:
\begin{eqnarray}
S_I &=& -\frac{1}{M^2}\int \{\delta^{b}_{c} \bar{l}_{\omega b i} R^{\omega a i}
+ \eta^{b}_{c} \bar{u}_{R,\omega,b} U^{\omega a}_L  +
\lambda^{b}_{c}\bar{d}_{R,\omega,b} D^{\omega a}_L\nonumber\\&&+\delta^{b}_{c}
\bar{l}_{ b i} R^{ a i} + \zeta^{b}_{c}\bar{\nu}_{R,b} N^{ a}_L
+\gamma^{b}_{c}\bar{e}_{R,b} E^{ a}_L\}^+\nonumber\\&&\{\delta^{b}_{c}
\bar{l}_{\omega b i} R^{\omega a i} + \eta^{b}_{c} \bar{u}_{R,\omega,b}
U^{\omega a}_L  + \lambda^{b}_{c}\bar{d}_{R,\omega,b} D^{\omega
a}_L\nonumber\\&&+\delta^{b}_{c} \bar{l}_{ b i} R^{ a i} +
\zeta^{b}_{c}\bar{\nu}_{R,b} N^{ a}_L +\gamma^{b}_{c}\bar{e}_{R,b} E^{
a}_L\}d^4x\label{FIN}
\end{eqnarray}

After Fierz rearrangement in a similar way to the previous section we obtain
mass matrices for the SM fermions:
\begin{eqnarray}
m_u &=& -\frac{1}{12 M^2}<\bar{U} U > \eta \nonumber\\
m_d &=& -\frac{1}{12 M^2}<\bar{D} D > \lambda\nonumber\\
m_{\nu} &=& -\frac{1}{4 M^2}<\bar{N} N > \zeta\nonumber\\
m_{e} &=& -\frac{1}{4 M^2}<\bar{E} E > \gamma
\end{eqnarray}

We can diagonalize these matrices via an appropriate unitary transformation of
leptons and quarks of different generations. During this diagonalization mixing
in the charged currents appears as usual.

\section{Conclusions}

In this paper we considered the alternative (to ETC) scheme of SM mass
generation for the Farhi-Susskind technicolor model. Additional $SU(N_{TC})$
scalar fields are introduced with masses well above Technicolor scale. Due to
coupling of these scalars to SM fermions and technifermions the transition
between both kinds of fermions appears. As a result SM fermions become massive.
Fermion mixing also appears in this scheme in a natural way. The action for the
mentioned scalar fields does not contain $\phi^4$ interaction. Therefore, the
given model does not suffer from the problems specific for the SM Higgs sector.

Probably, the main advance of the given approach is that all additional terms
in effective action (\ref{FIN}) contain transition between SM fermions and
technifermions. The additional four - fermion terms that contain only SM
particles do not appear. That's why a lot of difficulties due to appearance of
such terms in ETC models are avoided.

Nevertheless, several problems related to the suggested scheme remain unsolved.
In particular, the consideration of possible additional effects due to the
terms appeared in (\ref{FIN}) are in order. Also it is important to consider
carefully the question about the renormalizability of the given model and
quantum anomalies. All these issues, however, are considered to be out of the
scope of the present paper\footnote{When this work was completed the author
became aware that the idea to consider $SU(N_{TC})$ fundamental scalar fields
to couple SM fermions to technifermions was suggested in \cite{Kagan} within
supersymmetric technicolor model and was used later in a series of papers
\cite{Dobrescu_Kagan}}.

The author kindly acknowledges useful discussions with Yu.A.Simonov, which
stimulated him to consider the given problem. This work was partly supported by
RFBR grants 09-02-00338, 08-02-00661, by Grant for leading scientific schools
679.2008.2.

\clearpage

\end{document}